\documentclass[aps,prl,twocolumn,superscriptaddress,nolinenumbers,nobibnotes,nodoi,noeprint]{revtex4-2}
\usepackage{amsmath,amssymb,physics,bm,siunitx}
\usepackage{xcolor,graphicx}
\usepackage{soul}
\usepackage[colorlinks=true,citecolor=blue,urlcolor
=blue,linkcolor=black]{hyperref}

\begin{document}

\title{Quantum Coulomb drag mediated by cotunneling of fluxons and Cooper pairs}
\author{Andrew G. Semenov}
 \affiliation{I.E.Tamm Department of Theoretical Physics, P.N.Lebedev Physical Institute, 119991 Moscow, Russia} 
\affiliation{Skolkovo Institute of Science and Technology, 121205 Moscow, Russia}
\author{Alex Latyshev}
\affiliation{Departement de Physique Theorique, Universite de Geneve, CH-1211 Geneve 4, Switzerland}
\author{Andrei D. Zaikin}
\affiliation{I.E.Tamm Department of Theoretical Physics, P.N.Lebedev Physical Institute, 119991 Moscow, Russia} 
\affiliation{
National Research University Higher School of Economics, 101000 Moscow, Russia}

\date{\today}
\begin{abstract}
We predict two novel quantum drag effects which can occur in macroscopically quantum coherent Josephson circuits. We demonstrate that biasing one resistively shunted Josephson junction by an external current one can induce a non-zero voltage drop across another such junction capacitively coupled to the first one. This quantum Coulomb drag is caused by cotunneling of magnetic flux quanta across both junctions which remain in the "superconducting" regime. Likewise, Cooper pair cotunneling across a pair of connected in series Josephson junctions in the "insulating" regime is responsible for another -- dual -- quantum Coulomb drag effect.

\end{abstract}
\maketitle

{\it Introduction.} Josephson junctions \cite{BP} as well as other types of superconducting weak links \cite{KGI} exhibit a large variety of fundamentally important physical phenomena and -- at the same time -- provide a broad range of diverse applications. These systems offer a wonderful playground \cite{SZ90,book} enabling one to conveniently test quantum mechanics with dissipation on a macroscopic scale \cite{CL}.

Josephson junction circuits are being actively and successfully employed in order to realize superconducting qubits for quantum information processing \cite{qubits} giving rise to the field of circuit quantum electrodynamics \cite{CQED}. Quantum metrology is another possible important application of such structures. In particular, quantum phase slip junctions are treated as promising candidates for building the quantum standard of electric current \cite{curst,Bloch} by detecting the so-called Bloch steps dual to Shapiro ones routinely observed in "classical" Josephson junctions \cite{BP}.  Recently achieved synchronization of a pair of capacitively coupled ultrasmall Josephson junctions in the Bloch oscillations regime \cite{ptb} could serve as yet one more step on the way towards the quantum current standard.

Establishing macroscopic quantum coherence within a Josephson circuit in question is an important prerequisite for any of the above applications. In this Letter, we will demonstrate that quantum coherent behavior of coupled Josephson junctions gives rise to two novel -- dual to each other -- quantum drag effects. One of these effects is due to cotunneling of flux quanta through a pair of connected in parallel capacitively coupled junctions  in a configuration somewhat reminiscent of that studied in \cite{ptb} in a different physical limit. Another (dual) quantum drag effect is caused by cotunneling of Cooper pairs in a system of two connected in series Josephson junctions. In both cases by applying an external bias to one of the junctions one is able to control the quantum state of another one demonstrating a non-trivial behavior that we elaborate on further below.

{\it The model and effective action.} Consider the structure depicted in Fig. 1. This structure consists of two Josephson junctions (with capacitances $C_1$ and $C_2$ and Josephson coupling energies $E_{J1}$ and $E_{J2}$) shunted by Ohmic resistors $R_1$ and $R_2$ and connected via mutual capacitance $C_m$ as shown in the Figure. The first junction is biased by an external current $I$ whereas no electric current flows across the second junction. The voltage values across both junctions $V_1$ and $V_2$ are measured by voltmeters.
\begin{figure}[h]
   \centering
   \includegraphics[width=8cm]{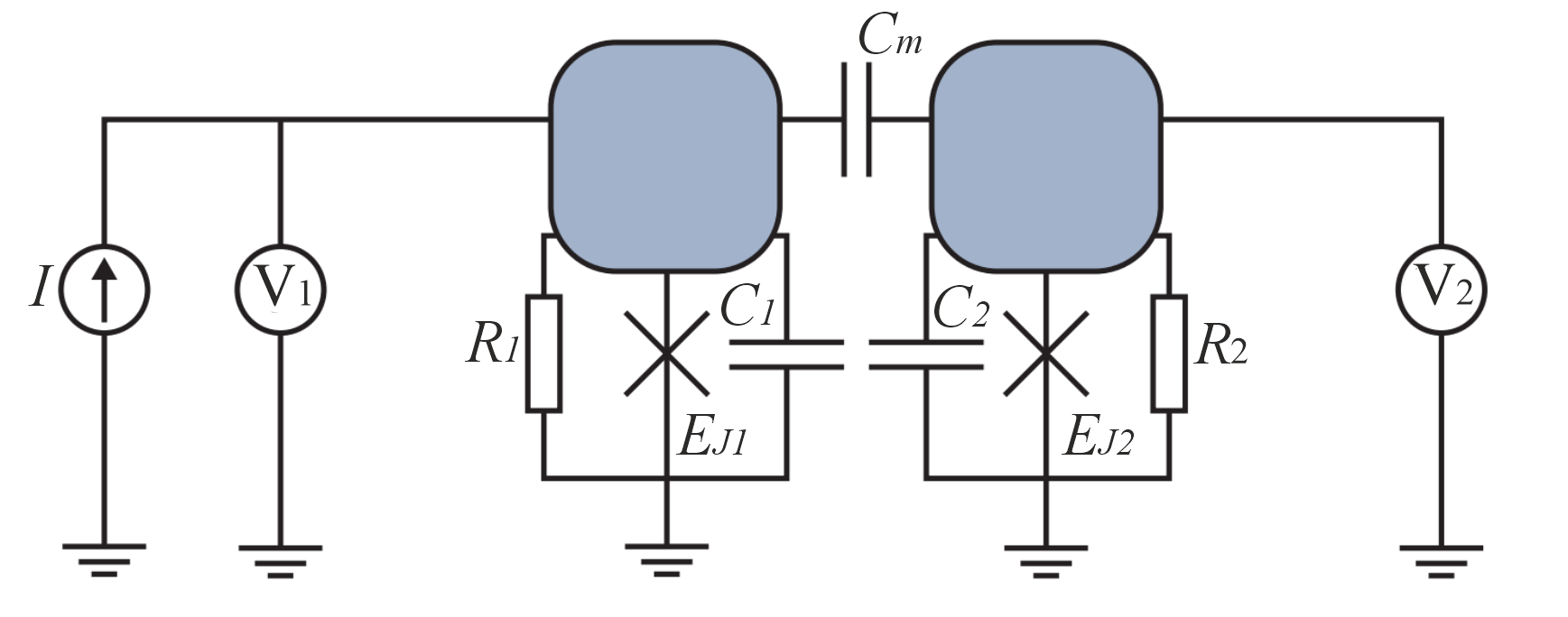}
{Fig. 1: The system I under consideration.}
\end{figure} 

The Hamiltonian for the structure in Fig. 1 can be written in the form
\begin{equation}
\hat{H}=\hat{H}_J+\hat{H}_{\rm env}+\hat{H}_{\rm int},
\label{Ham}
\end{equation}
where the first term in the right-hand side
\begin{gather}
\label{HamJ}
\hat{H}_J=\frac12\sum_{k,l=1,2}\hat{Q}_k\check{\mathcal C}^{-1}_{kl}\hat{Q}_l +U(\hat{\varphi}_1,\hat{\varphi}_2),\\
U(\hat{\varphi}_1,\hat{\varphi}_2)=\sum_{k=1,2}E_{Ji}(1-\cos \hat{\varphi}_k )-\frac{I\hat{\varphi}_1}{2e}
\label{U}
\end{gather}
defines the Hamiltonians of two Josephson junctions and the remaining two terms $\hat{H}_{\rm env}+\hat{H}_{\rm int}$ account respectively for the dissipative environment and for its interaction with the junctions degrees of freedom represented by the charge and the phase operators $\hat{Q}_{1,2}$ and $\hat{\varphi}_{1,2}$ obeying the standard commutation relations $[\hat{Q_k},\hat{\varphi}_l]=-i2e\delta_{kl}$, $k,l=1,2$. Note that here and below both Planck and Boltzmann constants $\hbar$ and $k_B$ as well as the speed of light $c$ are set equal to unity. The matrix elements $\check{\mathcal C}^{-1}_{kl}$ in Eq. \eqref{HamJ} are those of the inverted capacitance matrix
\begin{equation}
\check{\mathcal C}=
\begin{pmatrix}
C_{1} +C_m& -C_{m} \\
 -C_{m} & C_{2} +C_m\end{pmatrix}. 
\end{equation}
Here we do not need to specify the terms $\hat{H}_{\rm env}+\hat{H}_{\rm int}$ in Eq. \eqref{Ham} since all the environmental degrees of freedom are integrated out in the very beginning of the calculation according to the standard procedure \cite{SZ90,book}. As a result, one arrives at the grand partition function for our system expressed in terms of the double path integral
\begin{equation}
{\mathcal Z}=\tr \exp (-\hat{H}/T)=\int {\mathcal D}\varphi_1 \int {\mathcal D} \varphi_2 e^{-S[\varphi_1,\varphi_2]}.
\label{pathint}
\end{equation}
The effective action $S[\varphi_1,\varphi_2]$ reads
\begin{equation}
    S=S_1+S_2+\int_{-1/2T}^{1/2T} d\tau\left[\frac{C_m}{8e^2}\left(\frac{d(\varphi_1-\varphi_2)}{d \tau}\right)^2-\frac{I\varphi_1}{2e}\right],
\label{S}
\end{equation}
 where
\begin{gather}  
\label{action12} S_k=\int_{-1/2T}^{1/2T} d\tau \left[\frac{C_k}{8e^2}\left(\frac{d\varphi_k}{d\tau}\right)^2+E_{Jk}(1-\cos\varphi_k)\right]
\\+\frac{T^2R_Q}{8R_k}\int_{-1/2T}^{1/2T} d\tau \int_{-1/2T}^{1/2T} d\tau'\frac{[\varphi_k (\tau)-\varphi_k(\tau')]^2}{\sin^2 [\pi T(\tau-\tau')]}, \;\;k=1,2
\nonumber
\end{gather} 
and $R_Q=\pi /(2e^2) \simeq 6.45$ $k\Omega$ is the "superconducting" quantum resistance unit.

{\it Cotunneling of fluxons.} In order to proceed we will assume that both Josephson coupling energies strongly exceed all relevant charging energies of our problem, i.e. $E_{J1,2} \gg E_{C_k}=e^2/(2C_k)$, $k=1,2,m$. In this limit our Josephson "particle" with "coordinates" $\varphi_1$ and $\varphi_2$ is typically located close to one of the minima of the two-dimensional (2d) potential $U(\varphi_1,\varphi_2)$ \eqref{U} and may hop between the adjacent minima of this potential. 

For a moment let us set the external bias current equal to zero $I \to 0$. In the low temperature limit the only hopping mechanism is quantum tunneling. The tunneling amplitude $\gamma$ between different minima of $U(\varphi_1,\varphi_2)$ can be derived by evaluating the path integral \eqref{pathint} within the saddle point approximation. As usually, the corresponding saddle point trajectories are fixed by the equation $\delta S =0$ which reads
\begin{equation}
\int_{-1/2T}^{1/2T} d\tau' \check {\mathcal G}^{-1} (\tau-\tau') \left[\begin{array}{cc}
         \varphi_{1} (\tau') \\
         \varphi_{2} (\tau')   
          \end{array}\right]
    =\left[\begin{array}{cc}
        E_{J1}\sin\varphi_{1} (\tau) \\
         E_{J2}\sin\varphi_{2} (\tau)   
          \end{array}\right],
\label{Eq}
\end{equation}
where the matrix $\check {\mathcal G}^{-1}$ has the form 
\begin{equation}
\check {\mathcal G}^{-1}_{\omega_n}=\frac{1}{4e^2}[\check {\mathcal C}\omega_n^2+\check {\mathcal R}^{-1}|\omega_n|], \quad 
\check {\mathcal R}^{-1}=\begin{pmatrix}
\frac{1}{R_{1}} & 0 \\
 0 & \frac{1}{R_{2}}\end{pmatrix}
\label{G}
\end{equation}
and $\omega_n=2\pi nT$ (with all integer $n$) are the Matsubara frequencies. 

Provided one takes the limit $C_m \to 0$, Eqs. \eqref{Eq} decouple into two separate equations for $\varphi_1$ and $\varphi_2$. If, on top of that, one assumes that the inverse $RC$-times for both junctions remain much smaller than the corresponding plasma oscillation frequencies  
\begin{equation}
\Omega_k=\sqrt{8E_{Jk}E_{C_k}} \gg 1/(R_kC_k), \quad k=1,2,
\label{underdamped}
\end{equation}
dissipation can be neglected at this stage of our calculation by setting $\check {\mathcal R}^{-1} \to 0$ in Eq. \eqref{G}.  Then in the low temperature limit $T \to 0$ one immediately recovers the saddle point trajectories $\varphi_{1,2}=\pm \phi_{1,2}(\tau)$ describing quantum phase slippage process by $\pm 2\pi$ separately in each of the two Josephson junctions. Here  and below $\phi_{1,2}(\tau)=4\arctan(e^{\Omega_{1,2}\tau})$ is the well known instanton solution. Routinely taking into account fluctuations
around these instanton trajectories one arrives at the textbook result for the amplitude of quantum tunneling of the phases $\varphi_{1,2}$ between the adjacent potential minima 
\begin{equation}
\gamma_k=4\sqrt{\frac{E_{Jk}\Omega_k}{\pi}}\exp \left(-\frac{8E_{Jk}}{\Omega_k}\right), \quad k=1,2.
\label{amp0}
\end{equation}

\begin{figure}[h]
   \centering
   \includegraphics[width=8cm]{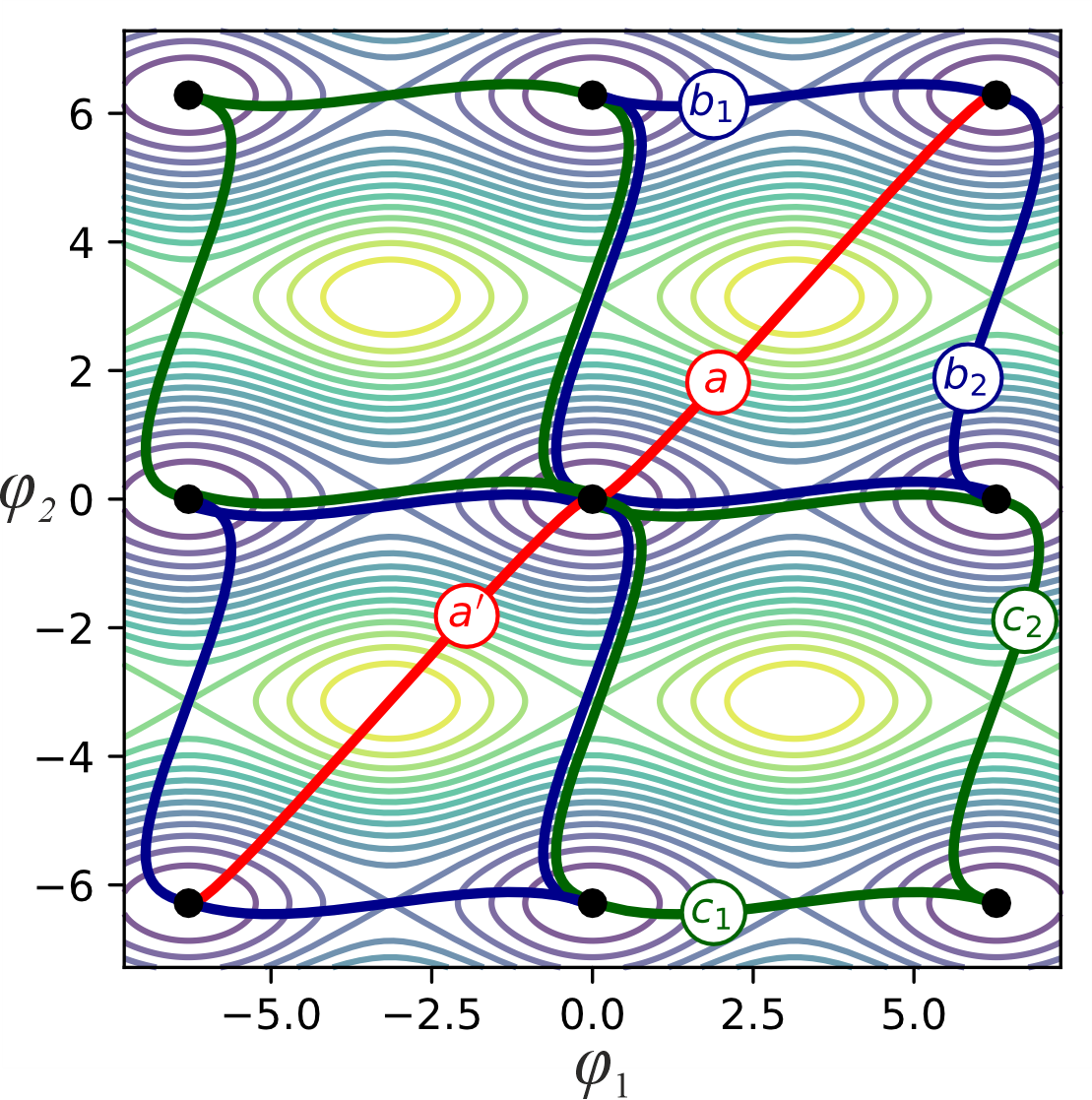}

   {Fig. 2: The paths  (a) (shown by red) and (b) (shown by blue) which account for tunneling between the states $(0,0)$ and $(2\pi,2\pi)$. The paths (c) (shown by green) correspond to tunneling between $(0,0)$ and $(2\pi,-2\pi)$. The paths pattern at $\varphi_1 <0$ is symmetric with respect to the origin.}
\end{figure}

The situation changes drastically for non-zero values of $C_m$. To begin with, the amplitude \eqref{amp0} now acquires the dependence on $C_m$, i.e. $\gamma_k \to \tilde\gamma_k \propto \exp (-A_0)$, where $A_0 \approx 8E_{Jk}/\tilde \Omega_k$ and $\tilde \Omega_k=\Omega_k \sqrt{1+C_m/C_k}$. Much more importantly, for $C_m>0$ quantum phase slips in different junctions interact with each other. Specifically, instantons with equal topological charges, e.g., $\phi_{1}(\tau)$ and $ \phi_{2}(\tau)$, {\it attract} each other and for not very small $C_m$ may glue together forming a new instanton that accounts for simultaneous tunneling of both phases between  the states (0,0) and $(2\pi, 2\pi )$ (cf. the path (a) in Fig. 2). At the same time, instantons with opposite topological charges,  e.g., $\phi_{1}(\tau)$ and $ -\phi_{2}(\tau)$, {\it repel} each other and, hence, cannot merge into one instanton. Accordingly, no direct tunneling between the states (0,0) and $(2\pi, -2\pi )$ would be possible. Of course, the phases at both junctions can still tunnel separately at different time moments via intermediate states (cf. the paths (b) and (c) in Fig. 2). However, such paths correspond to the action values bigger than those for the path (a) and, hence, their contributions can be neglected for sufficiently large $C_m$ (see below). Likewise, instantons $-\phi_{1}(\tau)$ and $ -\phi_{2}(\tau)$ can merge forming a single instanton trajectory (the path (a') in Fig. 2) that connects the states (0,0) and $(-2\pi, -2\pi )$ whereas no such single instanton trajectory exists for the states (0,0) and $(-2\pi, 2\pi )$. 

We can also add that each instanton corresponds to tunneling of one magnetic flux quantum $\Phi_0 = \pi/e$ (fluxon) across the Josephson junction in the direction perpendicular to the current. Merging of instantons at different junctions implies that tunneling of a fluxon across the first junction causes simultaneous tunneling of a fluxon in the second junction in the same direction. This is the effect of {\it cotunneling of fluxons} \cite{FN} through a pair of Josephson junctions displayed in Fig. 1.

In order to quantify our arguments let us evaluate the fluxon cotunneling amplitude $\gamma_{\rm cot}$ corresponding to the path (a) in Fig. 2. For simplicity below we will stick to the case of identical junctions with $E_{J1,2}=E_J$, $C_{1,2}=C$, $\Omega_{1,2}=\Omega=\sqrt{8E_JE_C}$ and  $\tilde\Omega_{1,2}=\tilde\Omega$. Provided the mutual capacitance $C_m$ remains sufficiently large attractive interaction between instantons at different junctions pushes them to effectively merge with each other. Under the substitution of $\varphi_1(\tau)=\varphi_2(\tau)=\varphi (\tau)$ into Eq. \eqref{Eq}, in the dissipativeless limit it reduces to
$\ddot\varphi=\Omega^2\sin\varphi$ with an obvious solution $\varphi (\tau)=\phi(\tau)=4\arctan(e^{\Omega\tau})$. After a straightforward calculation one finds
\begin{equation}
\gamma_{\rm cot} = B\exp (-A), \quad A = 16E_J/\Omega,
\label{A}
\end{equation}
where (see Supplemental material for details)
\begin{equation}
B=\sqrt{\frac{A}{2\pi}}\sqrt{\frac{\det[-\partial^2_\tau+\Omega^2]}{\det'[-\partial^2_\tau+\Omega^2\cos(\phi(\tau))]}}\sqrt{D}=\sqrt{\frac{32E_J\Omega D}{\pi}}.
\label{BB}
\end{equation}
Note that the product of the first two square roots in this equation just yields the standard result for a single junction with $C \to 2C$ and 
$E_J \to 2E_J$ (cf. Eq. \eqref{amp0}), whereas the term $\sqrt{D}$ with
\begin{equation}
   D=\frac{\det[-\partial^2_\tau+\omega^2]}{\det[ -\partial^2_\tau+\omega^2\cos(\phi(\tau))]}, \quad \frac{\omega^2}{\Omega^2}=\frac{1}{1+\frac{2C_m}{C}}
\label{D}
\end{equation}
accounts for the difference between two coupled Josephson junctions and a single one. Evaluating the ratio of the two determinants in Eq. \eqref{D} (see Supplemental material) one finds $D \to 1$ for $C_m \gg C$, i.e. in the strong interaction limit two junctions effectively behave as a single one. In the opposite limit $C_m \ll C$ we obtain  
\begin{equation}
B=8\sqrt{(3/\pi)E_J\Omega C/C_m}.
\label{Bpert}
\end{equation}
The latter result remains valid provided the action value on the path (a) equal to $A$ \eqref{A} exceeds that for two separate instantons (paths (b) and (c))  $\sim A\sqrt{1+C_m/C}$. Obviously, this is the case in the limit $C_m \gg C/A$. On the other hand, for $C_m \lesssim C/A$
the action values on the paths (a), (b) and (c) become practically equal and Eq. \eqref{Bpert} ceases to be valid. We also note that for $C_m >3.61C$ one has $\gamma_{\rm cot} > \tilde\gamma_{1,2}$, i.e. fluxon cotunneling becomes the dominant process.  

{\it Coulomb drag by fluxon cotunneling.} Let us now "turn on" both dissipation and the external current $I$ as shown in Fig. 1. We choose both shunting resistances such that $R_{1,2} < R_Q$, thus assuring that for $I \to 0$ and $T \to 0$ the Josephson "particle" remains localized in one of the potential minima \cite{SZ90,book,Schmid,Blg,Paco,FZ}. At $I>0$ the potential $U(\varphi_1,\varphi_2)$ \eqref{U} gets tilted in the "direction" of $\varphi_1$ and the "particle" starts sliding in this direction implying that a non-zero voltage proportional to the "particle velocity" $V_1=\langle \dot\varphi_1 /2e\rangle$ occurs across the first junction. 

Provided both temperature and the bias current values remain sufficiently low $T, I/e \ll \tilde\Omega$ the "particle" motion is due to its quantum tunneling between the adjacent potential minima, as we already discussed above. Each such tunneling event of a fluxon $\Phi_0$ implies a voltage pulse that temporarily occurs across the corresponding junction. Proceeding perturbatively in $\tilde\gamma_{1}$, in the lowest non-trivial order one finds \cite{SZ90,book}
\begin{equation}
V_1=\Phi_0(\Gamma_{1}(I)-\Gamma_1(-I)),
\label{V1} 
\end{equation}
where $\Gamma_{1}(I) \propto \tilde\gamma_1^2$ is the quantum decay rate of a metastable state for the phase variable $\varphi_1$ or, equivalently, the fluxon tunneling rate for the first junction. 

Very generally, the decay rate $\Gamma$ of any metastable state at low enough $T$ can be expressed in the form \cite{Langer,Weiss}
\begin{equation}
\Gamma =-2\Im F(T),
\label{ImF}
\end{equation}
where $F(T)= -T\ln {\mathcal Z}$ is the system free energy. The procedure that allows to establish the fluxon tunneling rate in a single Josephson junction is standard and is well documented in the literature (see, e.g., \cite{SZ90,book}). For instance, in order to find $\Gamma_{1}(I)$ one needs to consider the so-called bounce trajectory $\phi_{b}(\tau)$ that consists of two instantons with opposite topological charges
\begin{equation}
   \phi_{b}(\tau)= \phi (\tau-t_1)+\phi (t_2-\tau)-2\pi,\quad t_1<t_2.
\label{bounce}
\end{equation}
In the presence of Ohmic dissipation these two instantons attract each other logarithmically  and at $R_1 < R_Q$ form a close pair \cite{SZ90,book}. Substituting $\varphi_1(\tau)= \phi_{b}(\tau)$, $\varphi_2(\tau ) = 0$ into the action \eqref{S}, \eqref{action12} and integrating out the bounce zero mode one arrives at the lowest order correction to the free energy of the first junction $\delta F_1 \propto \tilde\gamma_1^2$ which formally diverges for any $I >0$. Performing an appropriate analytic continuation one recovers an imaginary part $\Im \delta F_1$ which -- being combined with Eq. \eqref{ImF} -- yields the expression for $\Gamma_1$. Then with the aid of \eqref{ImF} one arrives at the well known result \cite{SZ90,book}
\begin{equation}
\frac{V_1}{I} \sim \left(\frac{\tilde\gamma_1}{e\tilde\Omega}\right)^2\times \begin{cases}
(I/(e\tilde\Omega))^{2\alpha_1-2}, & I/e \gg T,
\\
(T/\tilde\Omega)^{2\alpha_1-2}, & I/e \ll T.
\end{cases}
\label{V1I}
\end{equation}
Here and below we define $\alpha_k=R_Q/R_k$, $k=1,2$.

The same procedure applied to the second junction yields $\Im\delta F_2=0$ since no current flows across this junction and, hence, no instability occurs. Accordingly, the tunneling rate $\Gamma_2 \propto \tilde\gamma_2^2$ vanishes and we have $V_2=0$.

The situation changes as soon as one gets to consider the process of fluxon cotunneling through both Josephson junctions. In this case we again deal with the bounce trajectory \eqref{bounce} but now we substitute $\varphi_1(\tau)=\varphi_2(\tau )= \phi_{b}(\tau)$ into the action \eqref{S}, \eqref{action12}. Proceeding in exactly the same way as above we arrive at the correction to the free energy $\delta F_{12}$ caused by fluxon cotunneling
\begin{equation}
\delta F_{12} = \gamma_{\rm cot}^2\int\limits_0^\infty dt e^ {I\Phi_0 t-2(\alpha_1+\alpha_2)\ln \left(\frac{|\sin(\pi Tt)|}{\pi T\Omega^{-1}}\right)}.
\label{F12}
\end{equation} 
Following the standard procedure \cite{SZ90,book,Weiss} one can deform the time integration contour in Eq. \eqref{F12} and continue the resulting expression analytically in order to recover the imaginary part $\Im F_{12}$. Then with the aid of Eq. \eqref{ImF} one recovers the fluxon cotunneling rate
\begin{eqnarray}
\nonumber
\Gamma_{\rm cot}(I)=\frac{\gamma_{\rm cot}^2}{\Omega}\left(\frac{\Omega}{2\pi T}\right)^{1-2(\alpha_1+\alpha_2)}\exp\left(\frac{I\Phi_0}{2T}\right)\\
\times\frac{|\varGamma (\alpha_1+\alpha_2+iI\Phi_0/(2\pi T))|^2}{\varGamma (2\alpha_1+2\alpha_2)}.
\label{Gcot}
\end{eqnarray}
As usually, detailed balance also yields
\begin{equation}
\Gamma_{\rm cot}(-I)=\exp \left(-\frac{I\Phi_0}{T}\right)\Gamma_{\rm cot}(I).
\label{dbr}
\end{equation}
This equation implies that a non-zero voltage $V_2=\langle \dot\varphi_2 /2e\rangle$ is generated across the second junction as long as a bias current $I>0$ is applied to the first junction. Similarly to Eq. \eqref{V1} one has
\begin{equation}
V_2=\Phi_0(\Gamma_{\rm cot}(I)-\Gamma_{\rm cot}(-I)).
\label{V2}
\end{equation}

Equations \eqref{Gcot}-\eqref{V2} combined with Eqs. \eqref{A}-\eqref{Bpert} for $\gamma_{\rm cot}$ account for Coulomb drag effect mediated by fluxon cotunneling and represent the first key result of this Letter. Remarkably, it turns out that in the configuration of Fig. 1 one can control the voltage across the second Josephson junction applying the current to the first one. In other words, due to fluxon cotunneling the Josephson "particle" in a tilted 2d washboard potential \eqref{U} propagates not only in the "direction" of a tilt (i.e. $\varphi_1$) but also along the $\varphi_2$-axis. The corresponding "velocity" follows directly from Eqs. \eqref{Gcot}-\eqref{V2} which yield 
\begin{equation}
\frac{V_2}{I} \sim \left(\frac{\gamma_{\rm cot}}{e\Omega}\right)^2\times \begin{cases}
(I/(e\Omega))^{2(\alpha_1+\alpha_2)-2}, & I/e \gg T,
\\
(T/\Omega)^{2(\alpha_1+\alpha_2)-2}, & I/e \ll T.
\end{cases}
\label{V2I}
\end{equation}

Note that cotunneling generates exactly the same correction of the form \eqref{V2I} also to the average voltage $V_1$ in the first junction. 
For $C_m <3.61C$ this correction is parametrically smaller than the leading order result \eqref{V1} and, hence, can be safely neglected. On the other hand, for $C_m >3.61C$ the cotunneling contribution \eqref{V2I} dominates over the one in Eq. \eqref{V1} and one has $V_1=V_2$.

{\it Phase-charge duality and Coulomb drag}. The above ideas can be developed further if one makes use of the well known duality property between the phase and the charge variables \cite{SZ90,book,ZP87,PZ88,averin,Z90}. In resistively shunted Josephson junctions this property allows to relate quantum dynamics of the charge to that of the phase by means of the duality transformation that includes 
\begin{equation}
\Phi_0 \leftrightarrow 2e, \;\;\; I \leftrightarrow V,\;\;\; \gamma_{1,2} \leftrightarrow \frac{E_{J1,2}}{2}, \;\;\; \alpha_{1,2} \leftrightarrow \frac{1}{\alpha_{1,2}},
\label{duality1}
\end{equation}
i.e. tunneling of Cooper pairs with charge $2e$ is dual to that of magnetic flux quanta $\Phi_0$.

\begin{figure}[h]
   \centering
   \includegraphics[width=8cm]{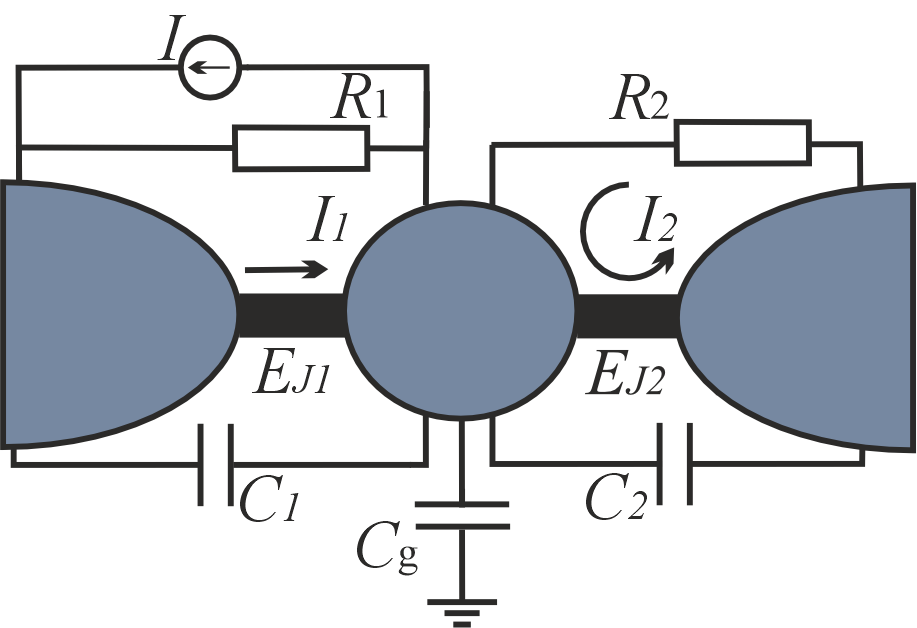}
   
{Fig. 3: The system II that exhibits the quantum drag effect dual to the one occuring in the system I displayed in Fig. 1.}
\end{figure} 

Consider now the system of two connected in series Josephson junctions (see Fig. 3) which are again described by the actions $S_1$ and $S_2$ \eqref{action12}. In the expression for the total action $S$ in Eq. \eqref{S} one should now set $C_m \to 0$ and add an extra term that accounts for the capacitance of the central island to the ground $C_g$. Proceeding in much the same way as in Ref. \onlinecite{LSZ} we immediately establish the correspondence between the fluxon cotunneling amplitude $\gamma_{\rm cot}$ and the Cooper pair cotunneling amplitude in the limit of small $E_{Jk} \ll \tilde E_C=e^2/2(C_1+C_2+C_g)$ and weak dissipation $\alpha_{1,2} \ll 1$:
\begin{equation}
\gamma_{\rm cot} \leftrightarrow E_{J1}E_{J2}/(4\tilde E_C).
\label{duality2}
\end{equation}
Note that in the above limit both junctions remain in the strong Coulomb blockade regime. Hence, the current $I$ biasing the first junction 
(apart from a small leakage current $I_1 \propto E_{J1}^2$ caused by weak tunneling of Cooper pairs \cite{SZ90,book,FZ,PZ88} through this junction) flows solely through the Ohmic shunt $R_1$ resulting in the average voltage drop $V \simeq IR_1$ across this junction.

Combining the duality conditions \eqref{duality1} and \eqref{duality2} with Eqs. \eqref{Gcot}-\eqref{V2I} we immediately recover the expression for the current $I_2$ flowing across another Josephson junction  
\begin{equation}
\frac{I_2}{V} \sim \left(\frac{eE_{J1}E_{J2}}{\tilde E_C/\tau_{RC}}\right)^2\times \begin{cases}
(eV\tau_{RC})^{\frac{2}{\alpha_1}+\frac{2}{\alpha_2}-2}, & eV \gg T,
\\
(T\tau_{RC})^{\frac{2}{\alpha_1}+\frac{2}{\alpha_2}-2}, & eV \ll T.
\end{cases}
\label{I2V}
\end{equation}
This result remains valid in the limit $eV, T \ll 1/\tau_{RC}$, where $1/\tau_{RC} \sim \tilde E_C \min (\alpha_1,\alpha_2)$ plays the role of an effective inverse $RC$-time. 

Equation \eqref{I2V} is the second key result of our work. It demonstrates that due to cotunneling of Cooper pairs one can control the average current flowing across one resistively shunted Josephson junction by effectively applying the average voltage $V$ to another one connected in series as shown in Fig. 3. Thus, we predict the second quantum Coulomb drag effect that is strictly dual to the one that occurs in the configuration displayed in Fig. 1.

In summary, we have predicted two quantum drag effects which can occur in different configurations of coupled resistively shunted ultrasmall Josephson junctions at low enough temperatures. Both these effects are mediated by the process of cotunneling. In the system displayed in Fig. 1 cotunneling of flux quanta $\Phi_0$ may lead to a non-zero voltage drop across one of the junctions induced by the current bias applied to another one. It is remarkable that in this case both junctions remain in the "superconducting" regime. Nevertheless one of them can sustain a non-zero average voltage even in the zero current limit. Cotunneling of Cooper pairs in the system of Fig. 3 leads to the second quantum drag effect that is dual to the first one. In this case even without any external bias one of the junctions can develop a non-vanishing current while being in the "insulator" regime. In order to realize this behavior it suffices to bias another junction which also remains in the "insulator" regime and is connected in series to the first one. These -- seemingly unusual -- properties result from the presence of macroscopic quantum coherence in our Josephson structures and can be directly tested in future experiments.

\begin{widetext}
\section{Supplemental material}
At sufficiently large values of $C_m$ the fluxon cotunneling amplitude is determined by the ratio of two path integrals
\begin{equation}
  \gamma_{\rm cot}= \frac{\int \mathcal D\varphi_1\mathcal D\varphi_2 e^{-\frac{C}{8e^2}\int d\tau \left(\dot\varphi_1^2+\dot\varphi_2^2+C_m(\dot\varphi_1-\dot\varphi_2)^2/C+\Omega^2\cos(\phi(\tau))\varphi_1^2+\Omega^2\cos(\phi(\tau))\varphi_2^2
   \right)} }{\int \mathcal D\varphi_1\mathcal D\varphi_2 e^{-\frac{C}{8e^2}\int d\tau \left(\dot\varphi_1^2+\dot\varphi_2^2+C_m(\dot\varphi_1-\dot\varphi_2)^2/C+\Omega^2\varphi_1^2+\Omega^2\varphi_2^2
   \right)}}.
\end{equation}
The integral in the numerator contains one zero mode which can be handled in a standard manner. Let us introduce two operators,
\begin{equation}
   \hat{\mathcal L}_0=\left(\begin{array}{cc}
      -(1+C_m/C)\partial^2_\tau+\Omega^2 & (C_m/C) \partial^2_\tau \\
      (C_m/C) \partial^2_\tau & -(1+C_m/C)\partial^2_\tau+\Omega^2
   \end{array}\right),
\label{L0}
\end{equation}
and
\begin{equation}
   \hat{\mathcal L}=\left(\begin{array}{cc}
      -(1+C_m/C)\partial^2_\tau+\Omega^2\cos(\phi(\tau)) & (C_m/C) \partial^2_\tau \\
      (C_m/C) \partial^2_\tau & -(1+C_m/C)\partial^2_\tau+\Omega^2\cos(\phi(\tau))
   \end{array}\right),
\label{L}
\end{equation}
respectively with the eigenvalues $\lambda_0^{(k)}$ and $\lambda^{(k)}$. One of the eigenvalues equals to zero $\lambda^{(0)}=0$ corresponding to the eigenfunction $\sim \dot\phi(\tau)$. This eigenvalue yields the contribution to the path integral equal to $T\sqrt{2\int d\tau \dot\phi^2(\tau)}$, whereas each non-zero eigenvalue contributes by the factor $\sqrt{8\pi e^2/(C\lambda^{(k)})}$. As a result, we arrive at Eq. (\ref{A}) for $\gamma_{\rm cot}$ with the pre-exponential factor $B$ defined as
\begin{equation}
   B=\sqrt{\frac{C}{4\pi e^2}\int d\tau \dot\phi^2(\tau)}\sqrt{\frac{\det[\hat{\mathcal L}_0]}{\det'[\hat{\mathcal L}]}},
\end{equation}
where the determinant in the denominator should be evaluated with the excluded zero mode.

Let us perform the orthogonal transformation of the operators \eqref{L0} and \eqref{L} with the aid of the matrix
\begin{equation}
   \hat O=\frac{1}{\sqrt{2}}\left(\begin{array}{cc}
      1 & 1 \\
      -1 & 1
   \end{array}\right).
\end{equation}
Then we obtain
\begin{equation}
   \hat O^T\hat{\mathcal L}_0\hat O=\left(\begin{array}{cc}
      -(1+2C_m/C)\partial^2_\tau+\Omega^2 & 0 \\
      0 & -\partial^2_\tau+\Omega^2
   \end{array}\right),
\end{equation}
\begin{equation}
   \hat O^T\hat{\mathcal L}\hat O=\left(\begin{array}{cc}
      -(1+2C_m/C)\partial^2_\tau+\Omega^2\cos(\phi(\tau)) & 0 \\
      0 & -\partial^2_\tau+\Omega^2\cos(\phi(\tau))
   \end{array}\right)
\end{equation}
and arrive at Eq. (\ref{BB}). The ratio of the first two determinants is known to be equal to
\begin{equation}
   \frac{\det[-\partial^2_\tau+\Omega^2]}{\det'[ -\partial^2_\tau+\Omega^2\cos(\phi(\tau))]}=4\Omega^2.
\end{equation}
The task at hand is to evaluate the remaining ratio of the two determinants $D$ in Eq. (\ref{D}). This task can be accomplished by means of Gelfand-Yaglom theorem. Let us introduce two solutions
\begin{equation}
   -\ddot \psi_{\lessgtr }(\tau)+\omega^2\cos(\phi(\tau))\psi_{\lessgtr }(\tau)=0,
\end{equation}
where
\begin{equation}
   \psi_{<}(\tau)\xrightarrow[\tau\to-\infty]{} e^{\omega\tau},\qquad\qquad \psi_{>}(\tau)\xrightarrow[\tau\to \infty]{} e^{-\omega\tau}.
\end{equation}
Bearing in mind that the imaginary time changes within the interval from $-1/(2T)$ to $1/(2T)$, we may construct the Gelfand-Yaglom solution in the form
\begin{equation}
   \psi(\tau)=\frac{\psi_<(\tau)\psi_>(-1/(2T))-\psi_>(\tau)\psi_<(-1/(2T)}{\dot\psi_<(-1/(2T))\psi_>(-1/(2T))-\dot\psi_>(-1/(2T))\psi_<(-1/(2T))}.
\end{equation}
Hence, the ratio of the determinants reads
\begin{equation}
   \frac{\det[-\partial^2_\tau+\omega^2]}{\det[ -\partial^2_\tau+\omega^2\cos(\phi(\tau))]}=\frac{\sinh(\omega / T)(\dot\psi_<(-1/(2T))\psi_>(-1/(2T))-\dot\psi_>(-1/(2T))\psi_<(-1/(2T)))}{\omega(\psi_<(1/(2T))\psi_>(-1/(2T))-\psi_>(1/(2T))\psi_<(-1/(2T)))}
\end{equation}
and $\psi_{<}(\tau)\xrightarrow[\tau\to \infty]{} e^{\omega\tau}/D$, $\psi_{>}(\tau)\xrightarrow[\tau\to -\infty]{} e^{-\omega\tau}/D$ with $D$ being defined in Eq. (\ref{D}).

In our particular case we have
\begin{equation}
   -\ddot\psi_{<}(\tau)+\omega^2\psi_{<}(\tau)-\frac{2\omega^2}{\cosh^2(\Omega\tau)}\psi_{<}(\tau)=0.
\end{equation}
Substituting $\psi_{<}(\tau)=u(\tanh(\Omega\tau))$ we transform this equation to the following Legendre equation
\begin{equation}
   (1-z^2)u''(z)-2zu'(z)-\frac{\omega^2}{\Omega^2(1-z^2)}u(z)+\frac{2\omega^2}{\Omega^2}u(z)=0.
\end{equation}
We choose the solution obeying all necessary boundary conditions. It reads
\begin{equation}
   P^{-\mu}_\nu(-z)=\frac{1}{\varGamma(1+\mu)}\left(\frac{1-z}{1+z}\right)^{-\mu/2} F\left(-\nu,\nu+1;1+\mu; \frac{1+z}{2}\right),
\end{equation}
where
\begin{equation}
   \mu=\frac{\omega}{\Omega}<1,\qquad\qquad \nu(\nu+1)=\frac{2\omega^2}{\Omega^2},\qquad\qquad \nu=\frac12\sqrt{1+\frac{8\omega^2}{\Omega^2}}-\frac12
\end{equation}
and $F(a,b,c,d)$ is the hypergeometric function.  Then we obtain
\begin{equation}
   \psi_{<}(\tau)=e^{\omega\tau}F\left(\frac12-\frac12\sqrt{1+\frac{8\omega^2}{\Omega^2}},\frac12+\frac12\sqrt{1+\frac{8\omega^2}{\Omega^2}};1+\mu; \frac{e^{\Omega\tau}}{2\cosh(\Omega\tau)}\right)
\end{equation}
and
\begin{equation}
   \frac{1}{D}=F\left(\frac12-\frac12\sqrt{1+\frac{8\omega^2}{\Omega^2}},\frac12+\frac12\sqrt{1+\frac{8\omega^2}{\Omega^2}};1+\frac{\omega}{\Omega}; 1\right).
\end{equation}
The last equation is equivalent to
\begin{equation}
D=\frac{\varGamma\left(\frac12+\frac{\omega}{\Omega}-\sqrt{\frac14+\frac{2\omega^2}{\Omega^2}}\right)\varGamma\left(\frac12+\frac{\omega}{\Omega}+\sqrt{\frac14+\frac{2\omega^2}{\Omega^2}}\right)}{\varGamma\left(1+\frac{\omega}{\Omega}\right)\varGamma\left(\frac{\omega}{\Omega}\right)},
\label{DD}
\end{equation}
where $\varGamma (x)$ is the Euler gamma-function. In the limit $C_m \ll C$ Eq. \eqref{DD} combined with Eq. (\ref{BB}) yields the result (\ref{Bpert}).
\end{widetext}


\begin{thebibliography}{99}
\bibitem{FN} Note that previously we have already discussed the effect of fluxon cotunneling \cite{LSZ} in a SQUID-like configuration caused by the mechanism essentially employing large inductive energy of a superconducting loop and being a full dual analogue of Cooper pair cotunneling in superconducting transistors. In contrast, the system analyzed here does not involve any inductive elements and the fluxon cotunneling mechanism is different from that analyzed in Ref. \onlinecite{LSZ}.
\end{thebibliography}
\end{document}